\documentclass[aps,prl,twocolumn,groupedaddress,showpacs]{revtex4}
\usepackage{color}
\usepackage{graphicx}
\usepackage{bm}

\usepackage{amsbsy}
\usepackage{amssymb}

\begin{document}
%\title{Stochastic nucleation of incommensurate clusters}
\title{Stochastic self-assembly of incommensurate clusters}

\author{M. R. D'Orsogna$^{1}$, G. Lakatos$^{2}$, and T. Chou$^{3}$}
\address{$^{1}$Dept. of Mathematics, CSUN, Los Angeles, CA 91330-8313\\
$^{2}$Dept. of Chemistry, The University of British Columbia,
Vancouver, BC, Canada, V6T-1Z1 \\
$^{3}$Depts. of Biomathematics and Mathematics, UCLA, Los Angeles, CA 90095-1766}
\date{\today}

%\runninglinenumbers

\begin{abstract}
%Nucleation, molecular aggregation, and growth are important processes
%in numerous physical and biological systems. In cellular biology,
%these processes often take place in confined spaces, involving a
%finite number of particles interacting in a stochastic manner.

We examine the classic problem of homogeneous nucleation and growth by
deriving and analyzing a fully discrete stochastic master equation.
Upon comparison with results obtained from the corresponding
mean-field Becker-D\"{o}ring equations we find striking differences
between the two corresponding equilibrium mean cluster
concentrations. These discrepancies depend primarily on the
divisibility of the total available mass by the maximum allowed
cluster size, and the remainder. When such mass incommensurability
arises, a single remainder particle can ``emulsify'' or ``disperse''
the system by significantly broadening the mean cluster size
distribution.  This finite-sized broadening effect is periodic in the
total mass of the system and can arise even when the system size is
asymptotically large, provided the ratio of the total mass to the
maximum cluster size is finite. For such finite ratios we show that
homogeneous nucleation in the limit of large, closed systems is not
accurately described by classical mean-field mass-action approaches.

%
%Recursion relations for determining metastable cluster concentrations
%in the coarsening regime are also derived.
%Upon enumeration of steady-state cluster configurations, we find
%recursion relations for determining quenched cluster configurations as
%well as exact analytical formulae for the equilibrium mean cluster
%size distribution.
%
%The configurational state-space spanned by the Master Equation is
%mapped onto a directed, rooted tree graph. In the limit of small but
%finite monomer detachment rates, all terminal vertices first become
%transiently populated, leading to a metastable mean cluster size
%distribution. If detachment is strictly forbidden, the cluster sizes
%are frozen to this distribution.  After the onset of monomer
%detachment and reattachment, the weighting shifts towards the highest
%order terminal vertices, corresponding to states with the smallest
%number of total clusters.  The corresponding {\it equilibrium} mean
%cluster size distribution can be determined from the relative weights
%among these highest order termini. The equilibrium cluster size
%distribution is shown to exhibit huge variations as a function of the
%total number of monomers, determined particularly by whether or no the
%total monomer number is an integer multiple of the maximum allowed
%cluster size.
\end{abstract}

\maketitle
\noindent
Nucleation and growth arise in countless physical and biological
settings \cite{KASH}.  In surface and material science, atoms and
molecules may nucleate to form islands and multiphase structures that
strongly affect overall material properties
\cite{AMAR1999}. Nucleation and growth are also ubiquitous in cellular
biology.  The polymerization of actin filaments \cite{SCIENCE} and
amyloid fibrils \cite{POWERS2006}, the assembly of virus capsids
\cite{ENDRES2002} and of antimicrobial peptides into transmembrane
pores \cite{BROGDEN2005}, the recruitment of transcription factors,
and the nucleation of clathrin-coated pits \cite{CLATHRIN} are all
important cell-level processes that can be cast as problems of
nucleation and growth for which there is great interest in developing
theoretical tools. Classical models of nucleation and growth
include mass-action kinetics, such as the Becker-D\"{o}ring (BD)
equations describing the evolution of the \textit{mean} concentrations
of clusters of a given size \cite{KASH}, or models of independent
clusters \cite{BARKEMA}. Solutions to the BD equations exhibit rich
behavior, including metastable particle distributions \cite{PENROSE},
multiple time scales \cite{WATTIS}, and nontrivial convergence to
equilibrium and coarsening \cite{JABIN, PENROSE}.  Within mean-field,
mass-action treatments however, correlations, discreteness or
stochastic effects are not included.  These may be important,
especially in applications to cell biology and nanotechnology, where
small system sizes or finite cluster ``stoichiometry'' are involved.

\begin{figure}[t]
\begin{center}
\includegraphics[width=3in]{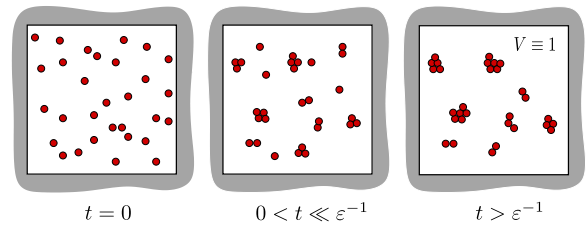}
\caption{(a) Homogeneous nucleation in a fixed, closed, unit volume
  initiated with $n_{1}(t=0)=M=30$ monomers. For small detachment
  rates monomers will be nearly exhausted at long times.  Here, the
  final cluster distribution consists of two dimers, one trimer, one
  4-mer, one pentamer, and two hexamers.}
\label{FIG1}
\end{center}
\end{figure}

In this paper, we carefully investigate the effects of discreteness
and stochasticity for a simple, mass-conserving homogeneous nucleation
process.  We construct the {\it probability} of the system to be in a
state with specified numbers of clusters of each size.  A
high-dimensional, fully stochastic master equation governing the
evolution of the state probabilities is derived, simulated, and solved
analytically in the equilibrium limit. Upon comparing the mean
cluster concentrations found from the stochastic master equation with
those obtained from the mean-field BD equations, we find qualitative
differences, even in the large system size limit.  Our results
highlight the importance of discreteness in nucleation and growth, and
how its inclusion leads to dramatically different results from those
obtained via classical, mean-field BD equations.

We begin by considering the simple homogeneous nucleation process in a
closed system (Fig.~\ref{FIG1}). Monomers first bind together to form
dimers.  Larger clusters are formed by successive monomer binding but
can also shrink by monomer detachment.  Within cellular
biophysics, nucleation and self-assembly often occur in small
volumes. Here, monomer production/degradation may be slow compared to
monomer attachment/detachment and the total number of monomers, both
free and within clusters, can be assumed constant.  Cluster sizes are
also typically limited, either by the finite total mass of the system,
or by some intrinsic stoichiometry. For example, virus
capsids, clathrin coated pits, and antimicrobial peptide pores
typically consist of $N\sim 100-1000, N\sim 10-20$, and $N\sim 5-8$
molecular subunits, respectively.  While various monomer binding and
unbinding rate structures \cite{PENROSE,WATTIS,JABIN,SMEREKA}, cluster
fragmentation/coagulation rules \cite{MAJUMDAR}, or the presence of
monomer sources \cite{WATTIS,BHATT} can be included, for the sake of
simplicity we consider only monomer binding and unbinding events
occurring at constant, cluster size-independent rates.  

%We focus on the strong binding limit since it best reveals the
%importance of stochasticity and finite-size effects.

Consider the probability density $P(\{n\};t)\equiv P(n_{1}, n_{2},
\ldots, n_{N}; t)$ of our system being in a state with $n_{1}$
monomers, $n_{2}$ dimers, $n_{3}$ trimers, $\ldots$, $n_{N}$
$N$-mers. The full stochastic master equation describing the time evolution of
$P(\{n\};t)$ is \cite{BHATT}
\begin{eqnarray}
\dot{P}(\{n\};t) & = &  -\Lambda(\{n\})P(\{n\};t) \nonumber \\
\: & \: & \displaystyle +{1\over
  2}(n_{1}+2)(n_{1}+1)W^{+}_{1}W^{+}_{1}W^{-}_{2}P(\{n\};t) \nonumber \\
\: & \: & \displaystyle +\varepsilon
(n_{2}+1)W^{+}_{2}W^{-}_{1}W^{-}_{1}P(\{n\};t) \nonumber \\
\: & \: &  \displaystyle  +
\sum_{i=2}^{N-1}(n_{1}+1)(n_{i}+1)W^{+}_{1}W^{+}_{i}W^{-}_{i+1}P(\{n\};t) \nonumber \\
\: & \: & \displaystyle  +\varepsilon
\sum_{i=3}^{N}(n_{i}+1)W^{-}_{1}W^{-}_{i-1}W^{+}_{i}P(\{n\};t).
\label{MASTERHOMO}
\end{eqnarray}
%\end{widetext}
\noindent
Here we non-dimensionalized time so that the binding rate is unity and
the detachment rate is $\varepsilon$. Since it best illustrates the
importance of discreteness in self-assembly, we henceforth restrict
ourselves to the strong binding limit $\varepsilon \ll 1$.  We define
$\Lambda(\{n\}) = {1\over 2}n_{1}(n_{1}-1) +
\sum_{i=2}^{N-1}\!n_{1}n_{i} + \varepsilon\sum_{i=2}^{N}n_{i}$ as the
total rate out of configuration $\{n\}$ and $W^{\pm}_{j}$ as the unit
raising/lowering operator that act the number of clusters of size $j$.
For example, $W^{+}_{1}W^{+}_{i}W^{-}_{i+1}P(\{n\};t) \equiv
P(\{n'\},t)$ where $\{n'\} =
(n_{1}+1,\ldots,n_{i}+1,n_{i+1}-1,\ldots).$ We assume that all the
mass is initially in the form of monomers: $P(\{n\};t=0) =
\delta_{n_{1},M}\delta_{n_{2},0}\cdots\delta_{n_{N},0}$.  By
construction, the stochastic dynamics described by
Eq.~\ref{MASTERHOMO} obey the total mass conservation constraint $M =
\sum_{k=1}^{N} k n_{k}$.

Solutions to Eq.~\ref{MASTERHOMO} can be used to define quantities
such as the mean numbers of clusters of size $k$: $\langle n_{k}(t)
\rangle \equiv \sum_{\{n\}} n_{k} P(\{n\};t)$.  These mean numbers
will be compared to the classical BD cluster concentrations $c_{k}(t)$
obtained by directly multiplying Eq.~\ref{MASTERHOMO} by $n_{k}$ and
summing over all allowable configurations. This procedure leads to a
hierarchy of equations relating the evolution of the mean $\langle
n_{k}(t) \rangle$ to higher moments such as $\langle
n_{j}(t)n_{k}(t)\rangle$. Closure of these equations using the
mean-field and large number approximations, $\langle n_k n_j \rangle
\simeq \langle n_k \rangle \langle n_j \rangle$ and $\langle n_1 (n_1
-1) \rangle \simeq \langle n_1 \rangle^2$, leads to the classical
Becker-D\"{o}ring equations
\begin{equation}
\begin{array}{rl}
\dot{c}_{1}(t) & = -c_{1}^{2}
- c_{1}\sum_{j=2}^{N-1}c_{j}  + 2\varepsilon c_{2} + \varepsilon \sum_{j=3}^{N}c_{j} \\[13pt]
\dot{c}_{2}(t) & = -c_{1}c_{2}+ {1\over 2}c_{1}^{2} - \varepsilon c_{2} + \varepsilon
c_{3} \\[13pt]
\dot{c}_{k}(t) & = -c_{1}c_{k}  + c_{1}c_{k-1} - \varepsilon c_{k} 
+\varepsilon c_{k+1} \\[13pt] 
\dot{c}_{N}(t) & = c_{1}c_{N-1} -\varepsilon c_{N},
\end{array}
\label{HOMOEQN3}
\end{equation}

\noindent where $c_{k}(t)$ is the mass-action approximation to
$\langle n_{k}(t)\rangle$. Here, the 
corresponding initial condition and
mass conservation are expressed as $c_k(t=0) = M \delta_{k,1}$ and $M
= \sum_{k=1}^{N} k c_k(t)$, respectively.  Eq.~\ref{HOMOEQN3} can be
easily integrated and analyzed at equilibrium 
in the $\varepsilon \ll 1$ limit

%By contrast, $c_{k}^{\rm eq}$ can be found by setting $\dot c_{k} = 0$
%in Eqs.\,\ref{HOMOEQN3} and perturbatively computing the resulting
%algebraic equations, so that
%
\begin{eqnarray*}
c_{k}^{\rm eq} \approx {\varepsilon \over 2} \left({2M \over \varepsilon N}\right)^{k/N}
\left[1- {k(N-1) \over N^{2}}\left({\varepsilon N\over 2M}\right)^{1/N} + \ldots\right].
%O\left((\varepsilon N)^{2/N}\right)\right].
\end{eqnarray*}
\noindent
where $c_k^{\rm eq} \equiv c_k(t \to \infty)$.  In equilibrium,
mean-field BD theory predicts maximal clusters of size $N$ dominate
with concentration $c^{\rm eq}_{N}\approx M/N$, while $c^{\rm
  eq}_{k<N} \sim \varepsilon^{1-k/N} \approx 0$ as $\varepsilon \to
0^{+}$. Under mass-action thus nearly all the mass is driven into the
largest cluster. However, a simple inconsistency emerges since the
solution $c_{k}^{\rm eq} \approx (M/N)\delta_{k,N}$ cannot be accurate
if $M<N$, when there is insufficient mass to form a single maximal
cluster.

\begin{figure*}[t]
\begin{center}
\includegraphics[width=5.6in]{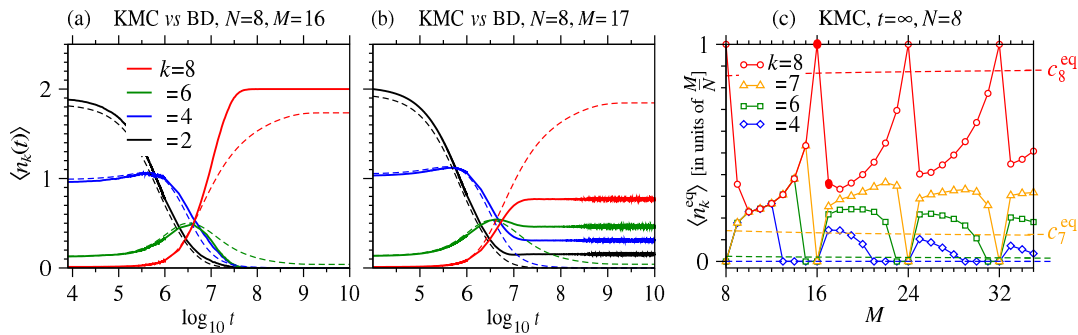}
\caption{Late-time mean cluster sizes $\langle n_{k}(t)\rangle$
  obtained from averaging $10^{6}$ KMC simulations of a stochastic
  nucleation process with $\varepsilon = 10^{-6}$. Only $k=2,4,6,8$
  are displayed.  (a) For $N=8$ and $M=16$, nearly all the mass is
  concentrated in $\langle n^{\rm eq}_{8}\rangle \approx 2$ at
  equilibrium. (b) For $N=8$, $M=17$, a much broader equilibrium mean
  cluster distribution arises. For comparison, the numerical solution
  for $c_{k}(t)$ from the BD equations are displayed by the dashed
  curves. The simulation and mean-field results agree well with each
  other, but a dramatic difference arises at long times as equilibrium
  is approached, particularly when the total mass $M$ is indivisible
  by $N$.
%Interestingly, even though the BD equations are most accurate
%  for short and intermediate times, it is during the intermediate
%  metastable regime that the variations in the cluster concentrations
%  are largest across simulation trajectories (see Supporting
% Information).  
(c) The difference between $c_{k}^{\rm eq}$ and $\langle n_{k}^{\rm
    eq}\rangle$ (plotted in units of $M/N$) is highlighted as a
  function of $M$.  The red dashed line corresponds to $c_{8}^{\rm
    eq}$ (which is nearly independent of $M$), while the open circles
  correspond to $\langle n_{8}^{\rm eq}\rangle$ found from Monte-Carlo
  simulation.  Note that $c_{8}^{\rm eq}\sim \langle n_{8}^{\rm
    eq}\rangle$ only when $M$ is divisible by $N=8$, or when $M/N \to
  \infty$.  The filled red circles correspond to $M=16$ and $M=17$ as
  detailed in (a) and (b), respectively. A few other mean
  concentrations, $\langle n_{4,6,7}^{\rm eq}\rangle$, along with the
  corresponding $c_{4,6,7}^{\rm eq}$ (dashed lines) are also plotted
  for reference.}
\label{MEAN}
\end{center}
\end{figure*}

To further investigate this inconsistency, we simulate the fully
stochastic Master equation (\ref{MASTERHOMO}) using a KMC or residence
time algorithm \cite{BKL75}. Figure \ref{MEAN} plots mean cluster
numbers $\langle n_{k}(t)\rangle$ and mean-field results $c_{k} (t)$
with $N=8$, $M=16,17$, and $\varepsilon = 10^{-6}$. Up to intermediate
times $t \lesssim \varepsilon^{-1}$, there is little difference
between the results for $M=16$ and $M=17$ and the mass-action
concentrations $c_{k}(t)$ roughly approximate $\langle
n_{k}(t)\rangle$. However, at long times $t\gg \varepsilon^{-1}$,
striking differences arise between the $M=2N=16$ and $M=2N+1 =17$
cases. We denote our solution in this limit as $\langle n_k^{\rm {eq}}
\rangle$, to be compared with $c_k^{\rm {eq}}$. For the commensurate
case $M=16$ (Fig.~\ref{MEAN}(a)) the mass-action solution $c_{k}^{\rm
  eq}$ roughly approximates $\langle n_{k}^{\rm eq}\rangle$, while for
the incommensurate case $M=17$, $c_{k}^{\rm eq}$ differs dramatically
from $\langle n_{k}^{\rm eq}\rangle$.  Figure \ref{MEAN}(c) highlights
the differences between $c_{k}^{\rm eq}$ and $\langle n_{k}^{\rm
  eq}\rangle$, particularly for $k=N=8$ (red curves). The
approximation $c_{k}^{\rm eq}\sim \langle n_{k}^{\rm eq}\rangle$ is
reasonable only when $M$ is exactly divisible by $N$, or, when $M$ is
very large.  In the latter case, the periodically-varying mean cluster
numbers $\langle n_{N}^{\rm eq}\rangle \to c_{N}^{\rm eq}$ as $M\to
\infty$, while all other $\langle n_{k<N}^{\rm eq}\rangle \to 0$.

\begin{figure}[t]
\begin{center}
\includegraphics[width=2.6in]{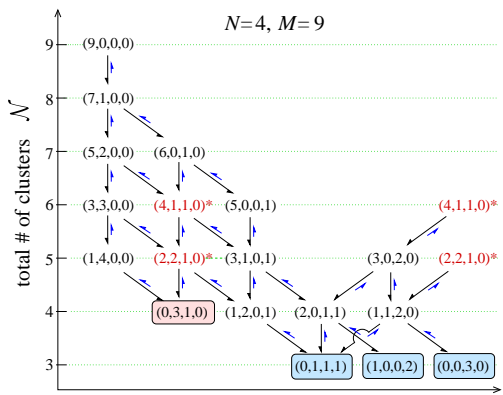}
\caption{Configurations $(n_{1}, n_{2}, n_{3},
  n_{4})$ for $N=4, M=9$.  Only three distinct states with a
  minimum number of clusters $\mathcal{N}_{\rm min} = 3$ arise. These
  are all connected by monomer attachment/detachment events to
  states with $\mathcal{N}_{\rm min}+1=4$ clusters. Applying
  detailed balance between them leads to 
  their weights in the $\varepsilon \to 0^{+}$ limit.  
%If $\varepsilon=0$, the system loses ergodicity and
%  appreciable probability will be forever trapped in state
%  $(0,3,1,0)$, leading to a very different metastable, pre-coarsening
%  distribution $\langle n_{k}^{*}\rangle$.
}
\label{STATE}
\end{center}
\end{figure}

To find analytic approximations to the equilibrium probabilities
$P(\{n\};t\to \infty)$, we make use of the fact that detachment is
slow.  In the $\varepsilon \ll 1$ limit, the most highly weighted
equilibrium configurations are those with the fewest total number of
clusters.  For each set $\{M, N\}$, we can thus enumerate the states
with the lowest number of clusters and use detailed balance to compute
their relative weights.  As an explicit example, consider the possible
states for the simple case $N=4, M=9$ shown in Fig.~\ref{STATE}. Here,
nearly all the weight settles into states with the lowest number of
clusters ($\mathcal{N}_{\rm min} = 3$ here).  Applying detailed
balance between the $\mathcal{N}_{\rm min} = 3$ and $\mathcal{N}_{\rm
  min}+1=4$ states, neglecting corrections of $O(\varepsilon)$, we
find $\langle n_{1}\rangle \approx \langle n_{2}\rangle \approx 6/13$,
$\langle n_{3}\rangle \approx 9/13$, and $\langle n_{4}\rangle \approx
18/13$.
%, which our KMC simulations confirm (see the Supporting
%Information for details).
%To extend our results to general $M$ and $N$, we start from the state
%with the highest possible number of maximum-sized clusters, given by
%the integer part of $[M/N]$, and distribute the remaining particles
%among smaller ones. The number of largest clusters is then
%successively reduced until all mass is exhausted. In this way, we
%inductively enumerate all states with near minimal total numbers of
%clusters.  We then use detailed balance to compute the relative
%equilibrium weights of these few-cluster states and find closed-form
%solutions for the mean equilibrium cluster numbers $\langle n_{k}^{\rm
%  eq}\rangle$.  
This process can be extended to general
$M$ and $N$ and leads to simple analytic solutions.
Upon defining $M=\sigma N - j$ where $\sigma$
denotes the maximum possible number of largest clusters, and $0 \leq j
\leq N-1$ represents the remainder of $M/N$, we arrive at one 
of our main results: exact solutions to the expected
equilibrium cluster numbers in the $\varepsilon\to 0^{+}$ limit:
\begin{eqnarray}
\langle n_N^{\rm eq} \rangle & = & 
\frac{\sigma (\sigma-1)}{(\sigma + j -1)}\label{EQ1}\\
\langle n_{N-k}^{\rm eq} \rangle & = & 
\frac{\sigma (\sigma -1) j (j-1) \dots (j-k+1)}
{(\sigma + j - 1)(\sigma+j-2) \dots (\sigma + j - k -1)}.\nonumber
\end{eqnarray}

\noindent These expressions are valid for $0\leq j < N-1$ and all $k$.
%Note that within all minimum cluster number states, the smallest
%cluster that can be formed is of size $N-j$. Therefore, $\langle
%n_{N-k}^{\rm eq}\rangle = 0$ for $k > j$, which is 
%satisfied by Eq.~\ref{EQ1}. 
In the special case $j = N-1$, the total mass can also be expressed as
$M=\sigma N -(N-1) = (\sigma-1)N + 1$ so that $j=N-1$ corresponds to
{\it adding} a single monomer to a system with $M=(\sigma-1)N$
monomers.  In this case, combinatoric factors of 2 that arise when
monomers appear in the populated configurations must be taken into
account leading to, for $j=N-1$,
\begin{eqnarray}
\langle n_1^{\rm eq} \rangle & = & \displaystyle  \frac {2
  (N-1)!} {D(\sigma,N-1)} \nonumber \\ 
\label{EQ2}
\displaystyle \langle n_{N-k}^{\rm
  eq} \rangle & = & \displaystyle 
              {\prod_{\ell=1}^{k}(N-\ell)\prod_{i=1}^{N-k-1}(\sigma-2+i)\over
                D(\sigma, N-1)} \\ \nonumber
%[j (j-1) \dots (j-k+1)] [(\sigma-1) \sigma \dots (\sigma+j-k-2)]
        \displaystyle \langle n_{N}^{\rm eq} \rangle & = & \displaystyle  (\sigma
        -1){D(\sigma -1, N-1)\over D(\sigma, N-1)},
\end{eqnarray}
\noindent where $D(\sigma,j) \equiv j! + \prod_{\ell=1}^{j-1}(\sigma +
\ell)$. Eqs.~\ref{EQ1} and \ref{EQ2} have been verified using
extensive Monte-Carlo simulations.  

\begin{figure}[t]
\begin{center}
\includegraphics[width=3in]{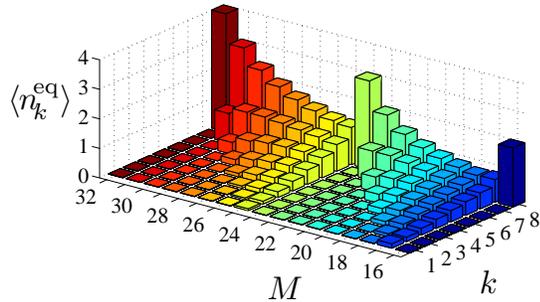}
\caption{The equilibrium cluster numbers $\langle n_{k}^{\rm
    eq}\rangle$ as $\varepsilon \to 0^{+}$, plotted as
  functions of $1\leq k \leq N=8$ and $M$.}
\label{NEQ}
\end{center}
\end{figure}

Fig.~\ref{NEQ} plots $\langle n_{k}^{\rm eq}\rangle$ for $N=8$ and
varying $M$.  Note that when $M=16,24,32$ is divisible by $N=8$ and
$j=0$, nearly all mass is deposited into the largest clusters, in
agreement with the mass-action BD results.  For cases where $M$ is not
an integer multiple of $N$ and $j > 0$, there are remaining monomers
that conspire to form smaller clusters. The number of ways this can
happen may be large, generating a broad distribution of cluster sizes.
For example, let us add a single monomer to the previously analyzed
(Fig.~\ref{MEAN}(a)) state $N=8,M=16$ ($\sigma = 2$, $j=0$). When
$M=17$ (Fig.~\ref{MEAN}(b)), Eq.~\ref{EQ2} can be used by setting
$\sigma =3$ and $j=N-1 = 7$.  Note that by adding just a
\textit{single} monomer, the mean cluster size distribution which for
$M=16$ was concentrated into the largest cluster, disperses and nearly
uniformly populates all cluster sizes. In our $N=8, M=17$ example,
$(1,0,0,0,0,0,0,2)$ is clearly one possible state with the lowest
number of clusters $\mathcal{N}_{\rm min} = 3$. However, as long as
some dissociation is allowed ($\varepsilon > 0$) a large number (in
this case $7$) of additional nontrivial 3-cluster states are possible:
$(0,1,0,0,0,0,1,1)$,$(0,0,0,1,0,1,1,0)$,$(0,0,1,0,0,1,0,1)$, 
$(0,0,0,0,2,0,1,0)$,$(0,0,0,1,1,0,0,1)$,$(0,0,1,0,0,0,2,0)$, $(0,0,0,0,1,2,0,0)$. The
equilibrium weights of these $8$ new states are comparable, resulting
in a very flat mean cluster size distribution, if compared to the
$N=8,M=16$ case.

We can quantify this ``dispersal'' effect by calculating the expected
cluster values $\langle n_k^{\rm {eq}} \rangle $ in the incommensurate
cases using Eqs.~\ref{EQ1} and \ref{EQ2}. As shown in Figs.~\ref{NEQ}
and \ref{MEAN}(c), when $M$ gets large, the dispersal effect
diminishes. Recall that the BD mass-action result $c_{k}^{\rm eq} \sim
(M/N)\delta_{k,N}$ puts all nearly all mass into $c^{\rm eq}_{N}$,
which is consistent with the exact solution in Eq.~\ref{EQ2} only when
$\langle n_{N-1}^{\rm eq}\rangle/\langle n_{N}^{\rm eq}\rangle \sim
N^{2}/M \ll 1$.  Thus, the mean-field result $c_{k}^{\rm eq} \sim
(M/N)\delta_{k,N}$ is asymptotically accurate only in the limit $M \gg
N^{2}$, or equivalently, when $\sigma \gg N$. Thus, the
periodically-varying curve $(N/M)\langle n_{k}^{\rm eq}\rangle$ in
Fig.~\ref{MEAN}(c) asymptotes to the mass-action result as $M/N^{2}
\to \infty$.

\begin{table}[tp]
\begin{tabular}{|p{3cm}|c|c|c|}
\toprule 
equilibrium cluster numbers ($\varepsilon \ll 1$) & \,\,${M\over
  N} \to 0$\,\, & \,\,${M\over N}$\, finite\,\,  & \,\,${M\over N} \gg N$\,\, \\[3pt]\hline
BD ($N=\infty$) & Eq. 2$^{*}$ & $\times$ & $\times$ \\[3pt]\hline BD
(finite $N$) & Eq. 2$^{*}$ & Eq. 2 &
Eq. 2$\dagger$ \\[3pt]\hline stochastic model & Eq. 6$^{*}$ &
\,\,Eqs. 6,7\,\, & \,\,Eqs. 6,7$\dagger$\,\, \\[3pt]\hline
\end{tabular}
\caption{Accuracy and validity regimes for equilibrium cluster numbers
  of different nucleation models for $\varepsilon \ll 1$. Results
  indicated by $^{*}$ or by $\dagger$ match in the $\varepsilon \to 0^{+}$
limit, but approach their common result very differently in $\varepsilon$.}
\label{TABLE}
\end{table}
% 

%It is in these incommensurate cases that the inadequacy of the
%classical BD equations becomes clear: here, BD results imply the
%predominance of the largest cluster size while our full discrete
%analysis shows the emergence of a much broader distribution.

%Our discrete stochastic nucleation model also allows us to consider
%nucleation under different particle number conditions and compare
%different models.  

Finally, Table \ref{TABLE} lists regimes of validity and results for
three different models: mass-action Becker-D\"{o}ring equations
without an imposed maximum cluster size, Becker-D\"{o}ring equations
with a fixed finite maximum cluster size $N$, and the fully stochastic
master equation. Three different ways of taking the large system
limits $M,N\to \infty$ are considered. The first column in Table
\ref{TABLE} with $N=\infty$ and $M$ finite corresponds to nucleation
with unbounded cluster sizes.  All models yield a single cluster of
size $M$, but display different scaling behavior in $\varepsilon$ (not
discussed here).  In the other extreme where $M/N \gg N$, equilibrium
results from the finite$-N$ BD equations match those of the discrete
stochastic model and all the mass is concentrated into clusters of
maximal size. However, just as before, the results from the
mass-action and stochastic treatments approach their common
distribution very differently in $\varepsilon$. The essential result
described in our work applies in the intermediate case where $M/N$ is
finite, as summarized in the middle column of Table \ref{TABLE}. Here
we find the novel incommensurability effect highlighted in
Figs.~\ref{MEAN}(c) and \ref{NEQ}.  These effects persist even in the
$M,N \to \infty$ limits, as long as their ratio is kept fixed.  Our
findings indicate that for many applications, where the effective
$M/N$ is finite, mean-field models of nucleation and growth fail and
discrete stochastic treatments are required.

%Ironically, in this
%regime, stochastic fluctuations are strongest 
%(see Supporting Information).

\vspace{4mm}

This work was supported by the NSF through grants DMS-1032131 (TC),
DMS-1021818 (TC), DMS-0719462 (MD), and DMS-1021850 (MD). TC is also
supported by the Army Research Office through grant 58386MA.

%%%%%%%%%%%%%%%%%%%%%%%%%%%%%%%%%%%%%%%%%%%%%%%%%%%%%%
%%%%%%%%%%%%%%%%%%%%%%%%%%%%%%%%%%%%%%%%%%%%%%%%%%%%%%
\end{document}